\documentclass[aps,pra,reprint,superscriptaddress]{revtex4-1}
\usepackage{graphicx}
\usepackage{dcolumn}
\usepackage{bm}
\usepackage{color}
\usepackage{calrsfs}

\usepackage{amsmath}
\usepackage{amssymb}
\usepackage{bbold}              
\usepackage{braket}
\usepackage[normalem]{ulem}     
\usepackage{csquotes}
\usepackage{listings}

\begin{document}

\title{Pulse-controlled qubit in semiconductor double quantum dots }

\author{Aleksander Lasek}
\email{alasek@umd.edu}
\thanks{equal contribution.}
    \affiliation{Cavendish Laboratory, J. J. Thomson Avenue, CB3 0HE, Cambridge, United Kingdom} 
    \affiliation{Hitachi Cambridge Laboratory, J. J. Thomson Avenue, CB3 0HE, Cambridge, United Kingdom}
\author{Hugo V. Lepage}
\email{hl407@cam.ac.uk}
\thanks{equal contribution.}
    \affiliation{Cavendish Laboratory, J. J. Thomson Avenue, CB3 0HE, Cambridge, United Kingdom}
\author{Kexin Zhang}
    \affiliation{Cavendish Laboratory, J. J. Thomson Avenue, CB3 0HE, Cambridge, United Kingdom}
\author{Thierry Ferrus}
    \affiliation{Hitachi Cambridge Laboratory, J. J. Thomson Avenue, CB3 0HE, Cambridge, United Kingdom}
\author{Crispin H. W. Barnes}
    \affiliation{Cavendish Laboratory, J. J. Thomson Avenue, CB3 0HE, Cambridge, United Kingdom}

\date{\today}

\begin{abstract}

We present a numerically-optimized multipulse framework for the quantum control of a single-electron charge qubit. Our framework defines a set of pulse sequences, necessary for the manipulation of the ideal qubit basis, that avoids errors associated with excitations outside the computational subspace. A novel control scheme manipulates the qubit adiabatically, while also retaining high speed and ability to perform a general single-qubit rotation.
This basis generates spatially localized logical qubit states, making readout straightforward.
We consider experimentally realistic semiconductor qubits with finite pulse rise and fall times and determine the fastest pulse sequence yielding the highest fidelity.
We show that our protocol leads to improved control of a qubit.
We present simulations of a double quantum dot in a semiconductor device to visualize and verify our protocol.
These results can be generalized to other physical systems since they depend only on pulse rise and fall times and the energy gap between the two lowest eigenstates.

\end{abstract}

\maketitle

\section{Introduction}

Accurate qubit control must maximize the probability that a qubit will remain in its computational basis.
In this work, we develop a numerically-optimized method for general control of a single qubit, accounting for finite control pulse rise time and potential imperfections. 
We demonstrate that in a double quantum dot (DQD) structure with two gates, where the pulses and the DQD potential are imperfect, a reliable qubit can still be defined and operated with high fidelity, using experimentally realistic parameters.
We achieve fast operations that are independent of the initial state, and do not induce excitations beyond the computational basis. 
While our framework is generic, we demonstrate its usefulness on semiconductor DQD charge qubits, owing to their usefulness and ubiquity. 
Semiconductor devices are attractive candidates for qubit hardware owing to their high compatibility with current industrial standards. 
They also benefit from decades in advances in processing and device integration that render processing costs low ~\citep{flamm2019}. 
Progress in fabrication and measurement techniques have led to extended coherence times and more precise and faster electronics, both for qubit control and readout, paving the way to scalability ~\cite{hayashi03,fujisawa06,petta05,petersson10,veldhorst14}. 
Within previous suggested architectures, DQDs offer a straightforward way of producing both charge and spin qubits ~\cite{lloyd93,divincenzo00}. 
For reading the qubit state, a charge detector \cite{Ferrus11} or even dispersive readout \citep{ciccarelli2011impedance,colless2013dispersive,gonzalez2015probing,crippa2017level} can be used. 
These detection methods are both achievable experimentally with great accuracy and speed owing to the improvement of charge detection sensitivity \cite{Fernando15}.

We thus use experimentally realistic parameters based on a semiconductor architecture. 
However, the results can be easily generalized to any qubit with a similar Hamiltonian form.


In this paper, we first model a effective potential for a generic DQD system to define the qubit basis states as bonding and anti-bonding states, in Section \ref{sec:methods}. In section \ref{sec:qubitcontrol}, we show how to initialize a single electron into one of the logical qubit basis states and how to perform a set of mutually orthogonal rotations on the Bloch sphere, thus an arbitrary rotation, using shaped pulses that correct for pulse rise time. We develop a pulsing scheme that is capable of generating time-optimized general unitary rotations despite imperfections, using only the voltage across the gate. We finally consider noise (section \ref{subsec:noise}), discuss our results (section \ref{sec:discussion}),  and give conclusions on the practicability of the scheme (section \ref{sec:conclusion}).

\section{Single-Electron Charge Qubit definition}\label{sec:methods}

It is generally assumed either for simplicity or ease of experimental manipulations that the logic basis-state wave functions of a DQD, $\ket{0}$ and $\ket{1}$, are fully localised in the left or right side of the DQD ~\cite{hayashi03,gorman05,dovzhenko11}. This assumption is convenient for \enquote{brute force} initialisation via applying a high bias voltage. The readout is also simple, realised by measuring the probability of the electron being in the left or right dot. However, in this case, quantum states necessarily contain contributions from higher energy eigenstates which give rise to additional composite oscillations, typically on timescales faster than the qubit oscillation itself~\cite{kataoka09}. They ultimately induce a loss of fidelity in gate operations. This issue is critical for practical implementations of quantum computation and schemes like bang-bang pulse sequences have been proposed in order to mitigate this effect \cite{bangbang}. Such sequences involve additional gate operations that could be detrimental to the overall operation time. Consequently, optimizing the qubit basis states is a necessary preliminary requirement before any other attempts at extending coherence or improving the gate fidelity.

If a linear combination of the two lowest eigenstates of the DQD system is used instead of assuming a fully localized state, a true two-level system is formed. A qubit control framework that doesn't involve energy states outside of the computational space would greatly improve the fidelity compared to the method above.

It is optimal to define the qubit states as equal combinations of the ground and first excited states at zero bias, because it produces well-localized qubits that can be measured while also preserving symmetry between the two logical states. This is demonstrated within the two-site localised state model ~\citep{jacek} (App. \ref{appendix:2siteLSM}). A zero-bias potential also makes the qubit first-order insensitive to electrical noise, improving fidelity \cite{kim2015}. Moreover, as described in section \ref{sec:qubitcontrol}, having zero detuning as a default achieves a high fidelity $R_{\vec{x}}$ rotation without any pulsing. 
The coefficients of the energy eigenstates must be equal in order to have symmetry between the qubit states. Therefore, for a given DQD potential $V_{\textup{DQD}}(x)$ (App. \ref{appendix:2siteLSM}), we define the logical states as:

\begin{align}
\begin{split}
\label{eq:OptimalQubit}
&\ket{0}  = \frac{\psi^{\textup{B}}(x) + \psi^{\textup{AB}}(x)}{\sqrt{2}}, \\ 
&\ket{1}  = \frac{\psi^{\textup{B}}(x) - \psi^{\textup{AB}}(x)}{\sqrt{2}},
\end{split}
\end{align}
where $\psi^{\textup{(A)B}}(x)$ is the (anti)bonding state wave function.

\begin{figure}
\centering
\includegraphics[width=1\columnwidth]{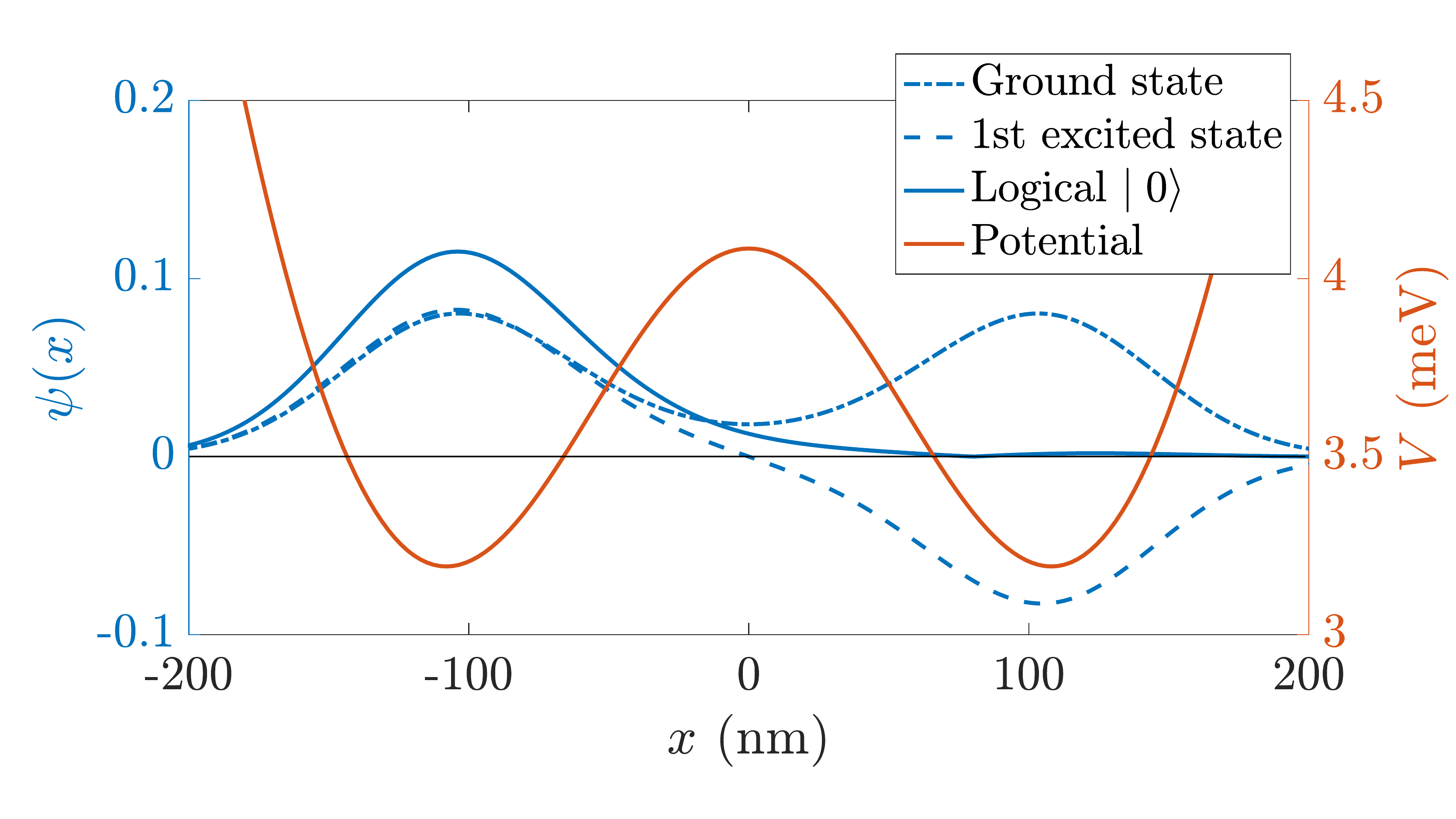}
\caption{Wave function of the two first excited states. The logical $\ket{0}$ and $\ket{1}$ qubits are formed using Eq. \ref{eq:OptimalQubit}. The values of the DQD spacing and the electrostatic potential amplitude were chosen for illustrative purposes and the scheme presented here works for a wide range of configurations.}
\label{fig:IdealQubit}
\end{figure}

While these states are not completely localized on a single dot, as their probability density is tailing to the other side of the dot (Fig.~\ref{fig:IdealQubit}), they do maximize the average probability of successful readout ~\citep{jacek}. Further localization of the states would introduce higher-energy states that would consequently not obey the ideal two-site Hamiltonian we aim to model (Eq.~\ref{eq:H2}). Since there is no reference to the underlying effective potential of the DQD in our definition, this qubit is well defined for potentials that are not symmetric and more generally, for any dot shape.

\section{Single Qubit Control}\label{sec:qubitcontrol}
In the energy eigenbasis, the Hamiltonian of the qubit system reads:

\begin{equation}\label{eq:H1}
\hat{H}(t) = -\frac{1}{2} \,\epsilon(t) \,\sigma_{x} + \frac{1}{2} \,\Delta \,\sigma_{z}  + \frac{1}{2}(E_{\textup{B}}+E_{\textup{AB}}).
\end{equation}

Here $E_{\textup{B}}$ and $E_{\textup{AB}}$ are the energies of the bonding and antibonding states, i.e. the two lowest energy states, at a linear detuning $\epsilon$ = 0 , $\Delta$ is the \enquote{hybridisation energy} between the two localised states, and $\sigma_{x/z}$ are the Pauli $x/z$ matrices.

Using the basis defined in Eq. \ref{eq:OptimalQubit}, where $\ket{0}$ and $\ket{1}$ are on the poles of a Bloch sphere, the Hamiltonian in Eq. \ref{eq:H1} is written as:

\begin{equation}\label{eq:H2}
\hat{H}_\mathrm{eff}(t) = -\frac{1}{2} \,\epsilon(t) \,\sigma_{z} + \frac{1}{2} \,\Delta \,\sigma_{x} .
\end{equation}
We have neglected the constant factor here. The time-dependent wave function can then be written in terms of the standard $\theta$ and $\phi$,  polar and azimuthal angles respectively, on the Bloch sphere:

\begin{equation}\label{eq:qubitAngles}
\psi(x,t) = \cos \! \left( \frac{\theta(t)}{2} \right) \ket{0} + e^{i \phi(t)} \sin \! \left (\frac{\theta(t)}{2} \right)\ket{1}.
\end{equation}

With no bias voltage, $\epsilon=0$, the wave function will undergo a constant rotation around the $z$-axis on the Bloch sphere. When applying a non-zero bias, the axis of rotation is shifted.

For the Hamiltonian in Eq.~\ref{eq:H2}, a general rotation on the Bloch sphere by an angle $\alpha$ around a direction $\vec{n}$ is given by the solution to the time-dependent Schr\"{o}dinger equation (TDSE):

\begin{equation}\label{eq:arbitraryRotation}
R_{\vec{n}} (\alpha(t)) = \mathcal{T} \exp \left( \frac{1}{i\hbar} \int_{0}^{t} \hat{H}_\mathrm{eff}(t') \mathrm{d} t' \right)
\end{equation}

where $\mathcal{T}$ is the time-ordering operator.

Rotations are performed by sending a bias voltage pulse of amplitude $V_{\mathrm{bias}} = \frac{\epsilon}{e \lambda}$ and duration $t_p$ to the double dot where $\epsilon$ and $\lambda$ are respectively the  detuning and voltage amplitude proportionality constant for a given potential. An instantaneous switch between the $V_{\mathrm{bias}} =0$ and $V_{\mathrm{bias}} =\frac{\epsilon}{e \lambda}$ bias states is generally preferred as this simplifies the dynamics and avoids spurious qubit rotations \cite{koppens06}. In this case, the detuning $\epsilon (t)$ is described as a set of step-functions and $R_{\vec{n}} (\alpha)$ is expressed analytically as a rotation of the qubit state around the axis on the Bloch sphere which passes through the eigenstates of $H (t')$ at a rate proportional to the difference in energy of these two eigenstates. Such a pulse requires a linear potential along the axis of the double dot, as in Eq.~\ref{eq:H1}, which is achieved by applying voltages to a set of metallic surface gates. 
During a square-wave pulse of detuning $\epsilon$, the system will evolve according to the Hamiltonian in Eq. \ref{eq:H2}, that will be constant during the on-time of the pulse, giving an unitary time evolution (rotation):
\begin{equation}\label{eq:arbitraryRotationUnitary}
U(t)=R_{\vec{n}}(\alpha(t)) = \exp \left( -i\frac{ \vec{n} \cdot \vec{\bf{\sigma}} }{2 \hbar}t\right),
\end{equation}
where $\vec{n}=(\Delta, 0 , \epsilon) $ is the axis of rotation, with rotation frequency given by its magnitude.

\begin{figure}
\centering
\includegraphics[width=1\columnwidth]{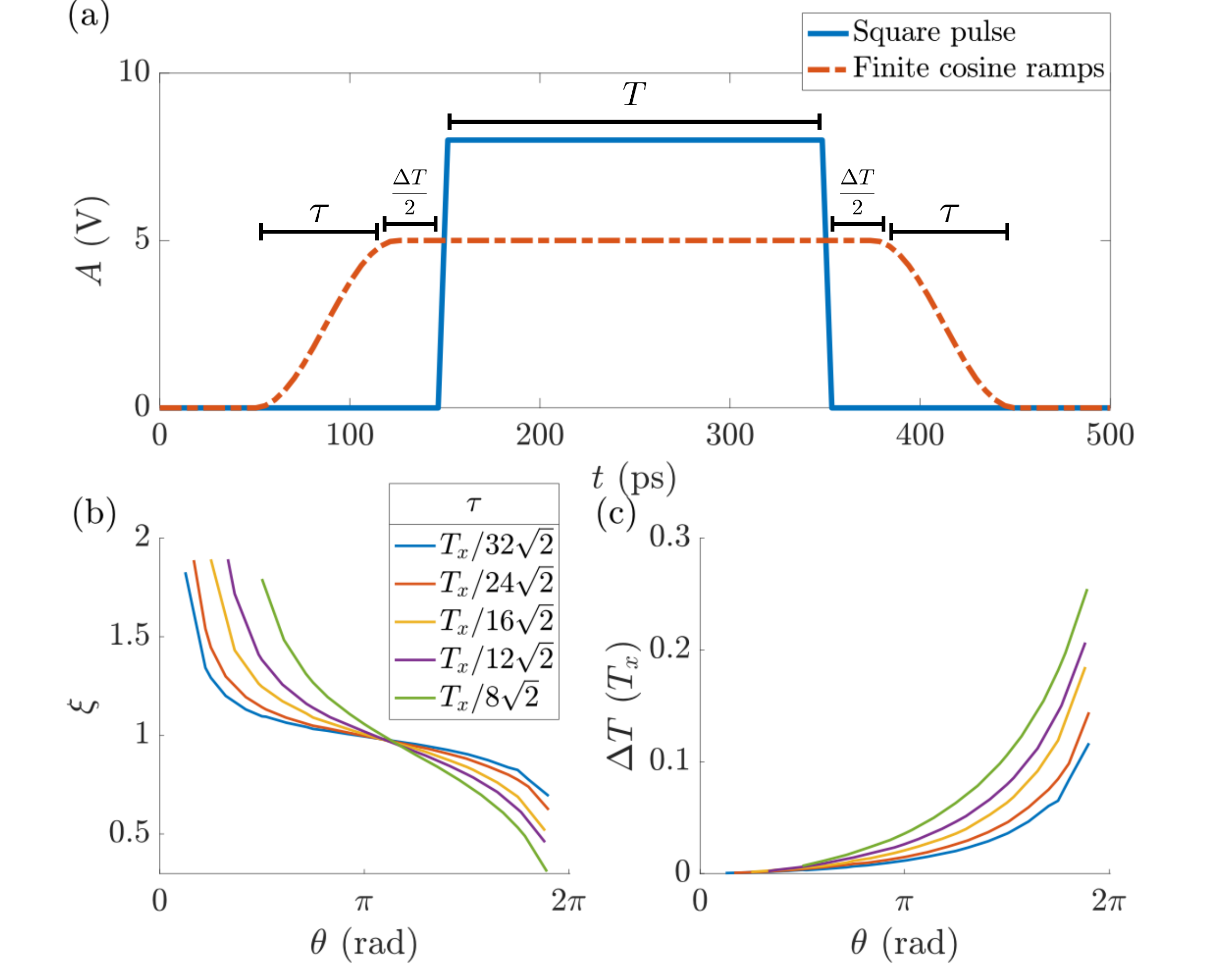}
\caption{\textbf{(a)} Amplitude profile of the ideal square pulse (solid blue line) to apply the linear bias given by Eq. \ref{eq:arbitraryRotationUnitary}. The pulse amplitude and duration are adjusted when the rise time $\tau$ is finite. Values of time (pulse duration, rise time) and voltage (pulse amplitude) are given for illustration purposes only. \textbf{(b)} Multiplicative amplitude adjustment factor $\xi$ given a target rotation angle $\theta$. Each coloured line corresponds to a different $\tau$ (see legend).  \textbf{(c)} Additive pulse duration adjustment $\Delta T$ with respect to the original square pulse time (see panel \textbf{(a)}). The rise times are not included in the additional pulse duration. Each coloured line corresponds to a different $\tau$ (see legend of \textbf{b}). }
\label{fig:PulseShape}
\end{figure}

Implementing such a square pulse isn't technically possible owing to practical limitations. Current and most commonly used pulse pattern generators have a built-in rise time $\tau$ of about 40ps to 500ps depending on the brand and characteristics. The Keysight 81134A Pulse Pattern Generator has a $\tau=$60ps between 20\% and 80\% of target amplitude. The Agilent 81130A and the Anritsu MP1763C have $\tau=$ 500ps and $\sim$ 40ps respectively, both between 10\% and 90\% of target amplitude. (Fig.~\ref{fig:PulseShape}a).

In this case, the step-function decomposition is not possible and, in general, Eq.~\ref{eq:arbitraryRotation} must be solved numerically. If the detuning can be described in terms of linear ramp functions, then Eq.~\ref{eq:arbitraryRotation} can be written analytically as a Landau-Zener-Stuckelberg transition \cite{landau32, zener32, stueckelberg32} but the resulting expression becomes a function of parabolic cylinder functions which makes understanding the rotation $R_{\vec{n}} (\alpha)$ more complex ~\citep{shevchenko2010, gradshteyn1994}.

In order to investigate the consequences that follow from this technical limitation, we have solved Eq.~\ref{eq:arbitraryRotation} numerically for a pulse with finite $\tau$ using a GPU-accelerated version of the staggered-leapfrog method~\cite{askar78, owen12, lepage2020, shukur2017, takada2019}  (see App.~\ref{appendix:ischeme}).

\begin{figure*}
\centering
\includegraphics[width=0.95\textwidth]{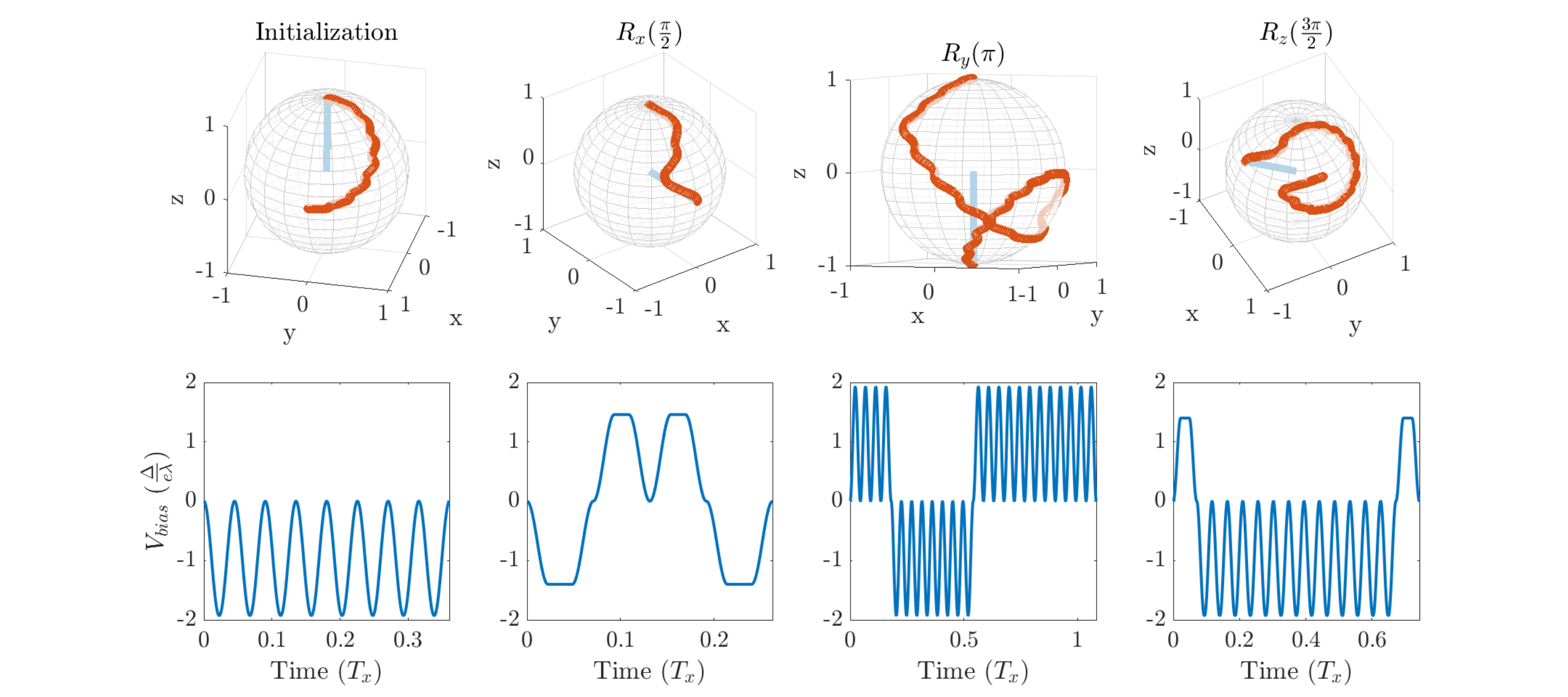}
\caption{Example pulse sequences and associated qubit rotations. Top: rotation path on the Bloch sphere. Bottom: Optimized pulse sequence where $T_x = \frac{2 \pi \hbar}{\Delta}$. To remain general, values of time and voltage are quoted as fractions of $T_x$ and $\frac{\Delta}{e\lambda}$ respectively. Exact experimental values will vary from one setup to the other. See the discussion for more details. All pulse sequences lead to a final state with a fidelity of $>$99.99\%. Furthermore, the same pulse sequence can be used for any initial state on the Bloch sphere (see section \ref{sec:fidelity}) without significant loss in fidelity.}
\label{fig:BlochSphere}
\end{figure*}

For such a pulse, the path of an individual qubit state on the Bloch sphere during the time evolution in Eq.~\ref{eq:arbitraryRotation} differs from the one induced by a square pulse \cite{foletti09} (Fig.~\ref{fig:BlochSphere}). In order to implement a high-fidelity rotation on the Bloch sphere, an effective  $R_{\vec{n}} (\alpha)$ is found by accounting for the aforementioned equipment limitations, such that the path traced on the Bloch sphere is different, but the resulting rotation remains the same as one induced by a perfect square pulse. We find that this can always be done by tuning the pulse duration and amplitude, depending on $\tau$ and desired angle of rotation. The details of this correction are outlined in \ref{subsec:correcting for rise time}. One can question whether such an adjusted operation including transient rotations is a proper rotation, i.e. independent of the initial state. The answer is yes, because while the precise path on the Bloch sphere may be difficult to describe analytically, the instantaneous Hamiltonian is still always expressed in terms of $\sigma_{x}$ and $\sigma_{z}$ matrices, therefore the effective operation is composed of rotations and is itself an actual rotation. We show that our pulses have the desired effect on any input state (section \ref{sec:fidelity}).
Additionally, it is worth noting that having a finite $\tau$ can have a desirable effect on the qubit, as it make the pulsing operation more adiabatic compared to using square pulses.

\subsection{General rotation scheme}\label{sec:general rotation scheme}
To perform an arbitrary qubit rotation, we propose a scheme of concatenating square pulses of alternating amplitudes. We set the bias voltage to produce a detuning $\epsilon=\pm \Delta$ , which gives the axes of rotation during pulsing to be in directions $(\frac{1}{\sqrt{2}},0,\pm \frac{1}{\sqrt{2}})$ on the Bloch sphere. We will call these axes $\vec{z}^{\prime}$ $(\frac{1}{\sqrt{2}},0, \frac{1}{\sqrt{2}})$ and $\vec{x}^{\prime}$ $(\frac{1}{\sqrt{2}},0,- \frac{1}{\sqrt{2}})$ respectively, as they are both rotated by $\frac{\pi}{4}$ around $\vec{y}$ w.r.t. the usual $\vec{z},\vec{x}$  axis of the Bloch sphere. An arbitrary rotation can be performed by combining up to five rotations around any two perpendicular axes, simply by aligning $\vec{x}^{\prime}$ with the desired axis of rotation $\vec{n}$, performing the rotation, and then reversing the first step. An arbitrary rotation by angle $\alpha$ can thus be performed around axis $\vec{n}$ in the following way:
\begin{align}
\begin{split}
\label{eq:rotation5}
    R_{\vec{n}} (\alpha) = & R_{\vec{x}^{\prime}}\left(\frac{\pi}{2}-\phi\right) 
    R_{\vec{z}^{\prime}}(\theta)
    R_{\vec{x}^{\prime}}(\alpha) \\
    & \cdot R_{\vec{z}^{\prime}}(-\theta)  
    R_{\vec{x}^{\prime}}\left[-\left(\frac{\pi}{2}-\phi\right)\right], 
\end{split}
\end{align}
where $\theta,\phi$ are the angles of $R_{\vec{y}}(\frac{\pi}{4}) \vec{n}$ on the Bloch sphere. 
The argument angles of the composite rotations correspond to durations of the composite pulses, with $2\pi$ corresponding to $T_{\mathrm{rot}}=\frac{2 \pi \hbar}{\sqrt{\Delta^2+\epsilon^2}}$, the period of a full rotation around $\vec{x}^{\prime}$ or $\vec{z}^{\prime}$ while bias voltage is on. Since the rotation around $\vec{x}^{\prime}$ or $\vec{z}^{\prime}$ is always in the positive direction, any negative angles have to be replaced by a positive complement of $2\pi$.

Eq. \ref{eq:rotation5} is simple to implement, but not optimal in operation time - it is known \cite{kaye2007} that three rotations are sufficient, which should result is a faster operation:

\begin{equation}
\label{eq:rotation3}
R_{\vec{n}} (\alpha) = e^{i \beta} R_{\vec{x}^{\prime}}(\Theta_1)
    R_{\vec{z}^{\prime}}(\Theta_2)
    R_{\vec{x}^{\prime}}(\Theta_3).
\end{equation}

Here, $\Theta_1$, $\Theta_2$ and $\Theta_3$ each depend on the angle and axis of the rotation.

\subsection{State preparation}\label{sec:state_preparation}

Before any quantum computation is performed, each qubit has to be initialized to a fiducial state, usually $\ket{0}$ or $\ket{1}$. For a generic operation involving a charge qubit, we would expect the initial state of the electron to be the ground state of the DQD (see Fig.~\ref{fig:IdealQubit}). Such a state is not part of the qubit's logical basis and an initial rotation is needed. In order to rotate the wave function from the ground energy eigenstate to the qubit $\ket{0}$ state, we can take advantage of knowing the initial state to simplify the operation. A $R_{\vec{z}^{\prime}}\left(\pi\right)$ rotation will initialise to the $\ket{0}$ state, while a $R_{\vec{x}^{\prime}}\left(\pi\right)$ will do so to the $\ket{1}$ state. Both are achieved with a single pulse, thus simplifying the initial state preparation.

\subsection{Single axis rotations}\label{subsec:sigmax}

Any single-qubit operation can be expressed in terms of rotations around two perpendicular axes. Here we provide the control sequence for rotations around the usual $\vec{x},\vec{y},\vec{z}$ Bloch sphere axes from an arbitrary point on the Bloch sphere.

The $R_{\vec{y}}$ rotation consists of only 3 pulses owing to angle cancellation in Eq. \ref{eq:rotation5} (as $\phi=\frac{\pi}{2}$):

\begin{equation}
\label{eq:rotateY}
    R_{\vec{y}} (\alpha) = R_{\vec{z}^{\prime}}\left(\frac{\pi}{2}\right) 
    R_{\vec{x}^{\prime}}\left(\alpha\right) 
    R_{\vec{z}^{\prime}}\left(\frac{3\pi}{2}\right) .
\end{equation}

To rotate in the opposite direction, one simply has to invert this pulse (swap $\vec{x}^{\prime}$ and $\vec{z}^{\prime}$) to get:

\begin{equation}
\label{eq:rotateYneg}
    R_{\vec{y}} (-\alpha) = R_{\vec{x}^{\prime}}\left(\frac{\pi}{2}\right) 
    R_{\vec{z}^{\prime}}\left(\alpha\right) 
    R_{\vec{x}^{\prime}}\left(\frac{3\pi}{2}\right) .
\end{equation}

$R_z$ and $R_x$ rotations would require five pulses if done as per Eq. \ref{eq:rotation5}. Instead, we solve Eq. \ref{eq:rotation3} for the angles to also perform them with just three pulses. Owing to symmetry, the first rotation is the same as the third one. Detailed derivation is presented in App. \ref{appendix:AnalyticalRotations}.
\begin{equation}
\label{eq:rotateXZ}
    R_{\vec{x}/\vec{z}} (\alpha) = R_{\vec{x}^{\prime}}\left(\Theta_1\right) 
    R_{\vec{z}^{\prime}}\left(\Theta_2\right)
    R_{\vec{x}^{\prime}}\left(\Theta_1\right),
\end{equation}

where
\begin{equation}
\Theta_1 = \arccos\left( \frac{\sqrt{2}\cos{\frac{\alpha}{2}}}{\sqrt{\cos{(\frac{\alpha}{2})}^2 +1}}\right),
\end{equation} 
and
\begin{equation}
\Theta_2 = 2\arctan\left({\sin{\Theta_1}}\right)
\end{equation} 

for $R_{\vec{x}}$, and

\begin{equation}
\Theta_2 = 2\left(\pi -\arctan\left({\sin{\Theta_1}}\right)\right)
\end{equation} 

for $R_{\vec{z}}$.
Additionally, we note that $-R_{\vec{z}}(\alpha) = R_{\vec{z}}(-\alpha)$, which allows us to shorten operation time for rotations with $\alpha \geq \frac{\pi}{2}$ by inverting the pulse profile to perform the complementary rotation instead.

We note that one of the effects of defining the qubit as in Eq. \ref{eq:OptimalQubit} is that $R_{\vec{x}}$ rotation will occur automatically due to the Hamiltonian, with the rotation period $T_x=\frac{2 \pi \hbar}{\Delta}$. In the many qubit case, all the qubits rotate at their respective frequencies, and one would usually work in the rotating basis, therefore an $R_{\vec{x}}$ rotation still needs to be performed as per Eq. \ref{eq:rotateXZ}.

Instead of using the usual $\vec{x},\vec{y},\vec{z}$ basis, we can instead use the $\vec{x}^{\prime},\vec{y},\vec{z}^{\prime}$ basis which is more natural for the detuned system, and can be used to define logic gates with fewer pulses. A single $R_{\vec{y}}(\frac{\pi}{4})$ rotation is required to move into this basis. $R_{\vec{x}^{\prime}},R_{\vec{z}^{\prime}}$ are then achieved with a single pulse, while $R_{\vec{y}}$ requires three, as in Eq.\ \ref{eq:rotateY}. This way, any computation can be performed in the rotated basis, where operations are quicker. At the end, one would need to rotate back to $\vec{x},\vec{y},\vec{z}$ using a $R_{\vec{y}}(-\frac{\pi}{4})$ rotation, for optimal readout of localised states.

Some logic gate examples are:

\begin{equation}
    X= R_{\vec{z}^{\prime}} (\pi) ,
\end{equation}
\begin{equation}
    Y= R_{\vec{y}} (\pi) ,
\end{equation}
\begin{equation}
    Z= R_{\vec{x}^{\prime}} (\pi) ,
\end{equation}
\begin{equation}
    H= R_{\vec{y}} (\frac{\pi}{2}) R_{\vec{x'}} (\pi) ,
\end{equation}
\begin{equation}
    R_{\phi}= R_{\vec{x}^{\prime}} (\phi) .
\end{equation}

\subsection{Correcting for rise time}
\label{subsec:correcting for rise time}
To account for the actual experimentally realisable pulses not being square due to rise time and limited bandwidth,
the bias voltage and pulse duration have to be adjusted. This adjustment depends on the target rotation angle and $\tau$, but not on the input state. Therefore, it is sufficient to optimize a single pulse for the instrument rise time and range of desired rotations - these single pulses can then be concatenated into three-pulse trains to achieve arbitrary qubit rotations of high fidelity.
Here we numerically find the correct adjustments. This allows experimentalists to apply the ideal control sequence by simply changing the amplitude and duration of each square pulse in the train, avoiding complicated pulse shapes while retaining high fidelity. 

We present the numerical results for required amplitude $\xi$ and pulse duration $\Delta T$ adjustments, depending on $\tau$ and angle of rotation $\alpha$, all expressed in terms of the physical system parameters. Here, $\xi$ is a multiplicative factor adjusting the amplitude with respect to the square pulse amplitude ($\xi = 1$), and $\Delta T$ is the additive time adjustment with respect to the square pulse duration as well, as per Fig. \ref{fig:PulseShape} (a) - it is always greater than zero. We use generalised rise times expressed in terms of a fraction of generalised time $T_x$ (period of a full rotation without any pulsing), as seen in the legend of Fig. \ref{fig:PulseShape}. We have chosen these values to correspond to minimum possible rotation angles of $\frac{\pi}{8}, \frac{\pi}{6}, \frac{\pi}{4}, \frac{\pi}{3}, \frac{\pi}{2}$, from shortest to longest. These are the minimum possible rotations, because they are given by a pulse that consists only of rising/falling time, with no flat top, and is therefore the shortest pulse of desired amplitude that is possible. Of course, it is still be possible to rotate by an arbitrarily small angle indirectly by adding a $2\pi$ rotation.
As can be seen in Fig. \ref{fig:PulseShape} (b,c) the required time adjustment rises exponentially with desired rotation angle. Therefore, it is optimal to compose any pulse of the smallest possible rotations, as this will result in shorter overall rotation time. If the target rotation angle does not subdivide into an integer number of shortest possible rotations, one needs to use somewhat longer sub-pulses appropriately. Assuming a sine-shaped rise ramp, this short pulse is a sine wave, which is straightforward to generate experimentally. Single qubit control can be achieved by sending sine waves, with frequency as high as experimentally possible, and amplitude given by $\xi$ in Fig. \ref{fig:PulseShape} (b). Note that the only system-specific quantity is the energy gap $\Delta$ - the signal frequency is independent of the qubit system and not resonant with the two-level system, and instead purely defined by experimental limitations of the equipment ($\tau > 0$). We present examples of rotations performed with this scheme in Fig. \ref{fig:BlochSphere}, which summarises our main results.

As the resulting fidelity varies significantly with even small deviations from the parameters found here, we find that trying to fit analytical expressions to the data is not very useful if high fidelity is required. While  $\Delta T$ as a function of rotation angle $\theta$ seems to be an exponential, while $\xi$ is a rotated S-curve, attempts to fit it results with unacceptably low fidelity for a large $\theta$ range. Therefore, we suggest the gradient ascent search procedure described here be performed for the system of interest, taking into account the specificity of the experimental setup. This could be done using numerical simulations like in this work, or directly by taking actual measurements in an experiment. However, the latter might not be practical, as we find that thousands of fidelity evaluations are necessary to find good enough adjustment parameter values. If significant measurement error is present, the required number of experimental runs necessary might not be possible to realise, further highlighting the need for numerical simulations. Pseudo-code of the gradient ascent procedure is provided in App. \ref{appendix:pseudocode}  - it should enable anyone to find the optimal parameters in a general case, for rise time and angles that are required.

\subsection{Fidelity as a function of initial state}
\label{sec:fidelity}

\begin{figure*}
    \centering
    \includegraphics[width=\textwidth]{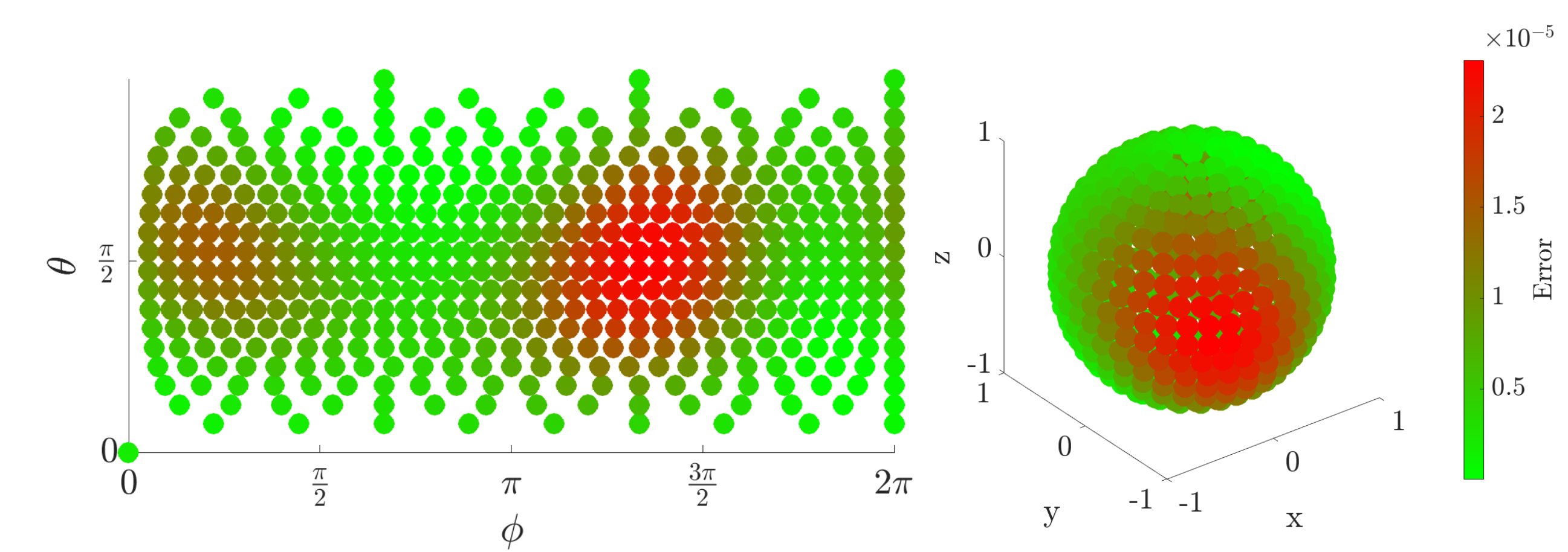}
    \caption{Error in fidelity for an $R_x(\pi)$ rotation as a projection (left) and on the surface of the Bloch Sphere (right).}
    \label{fig:Rx_pi}
\end{figure*}

\begin{figure*}
    \centering
    \includegraphics[width=\textwidth]{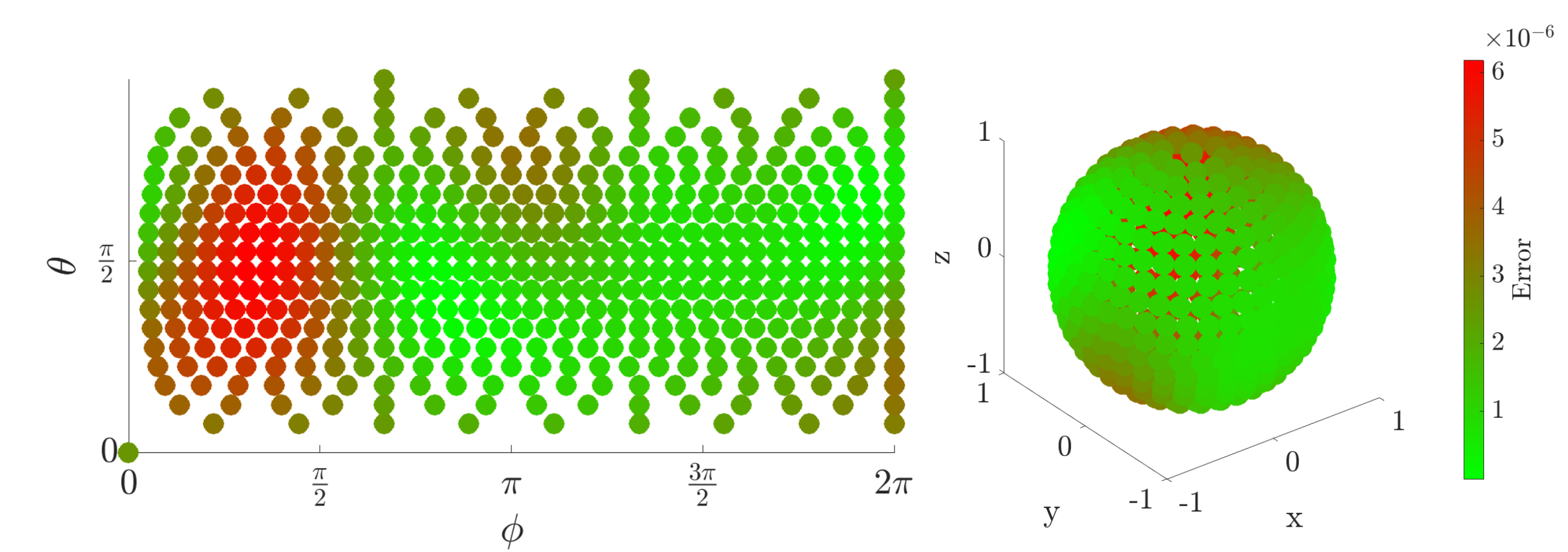}
    \caption{Error in fidelity for an $R_y(\pi)$ rotation as a projection (left) and on the surface of the Bloch Sphere (right).}
    \label{fig:Ry_pi}
\end{figure*}

\begin{figure*}
    \centering
    \includegraphics[width=\textwidth]{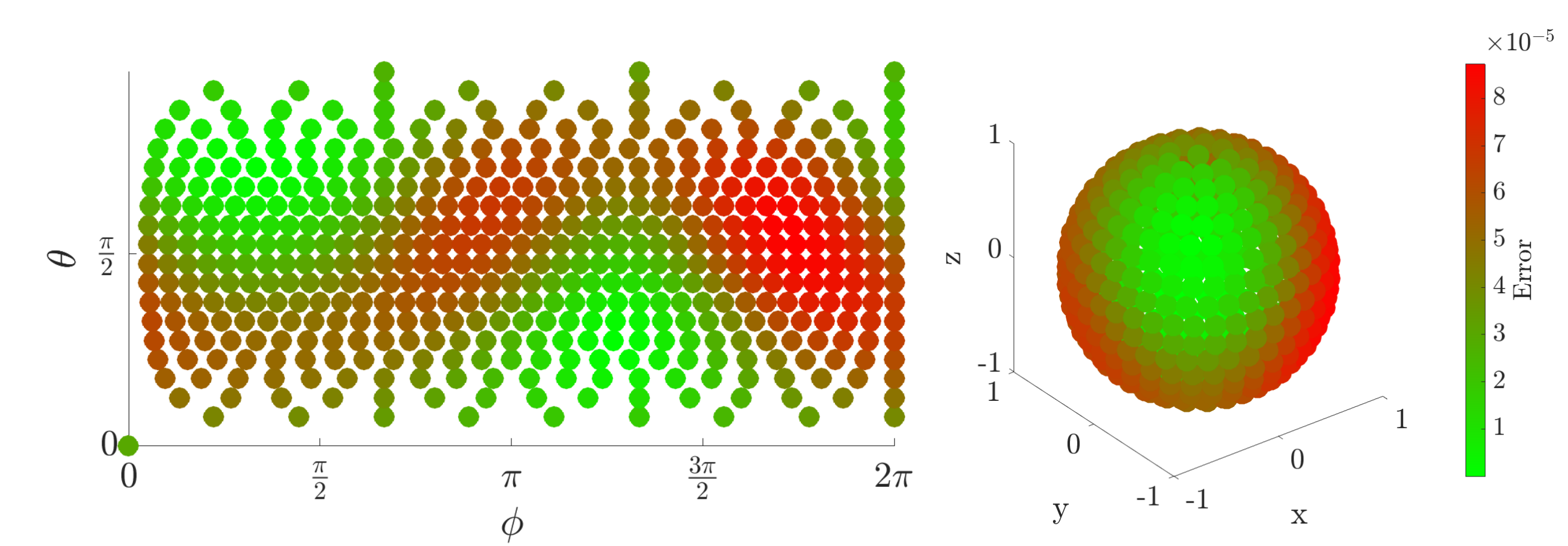}
    \caption{Error in fidelity for an $R_z(\pi)$ rotation as a projection (left) and on the surface of the Bloch Sphere (right).}
    \label{fig:Rz_pi}
\end{figure*}

The error in fidelity is found using 1-Fidelity, where fidelity is the overlap between the target state and the iterated state. Although some variation in fidelity is dependent on the initial state of the electron, any errors are below $10^{-4}$, and as low as $10^{-8}$ for some initial positions. This error could be reduced further if necessary by fine-tuning the adjustment parameters $\xi, \Delta T$. Figures \ref{fig:Rx_pi}, \ref{fig:Ry_pi}, and \ref{fig:Rz_pi} show a fidelity map for the $R_x$, $R_y$ and $R_z$ rotations respectively, as a function of Bloch sphere angles $\theta, \phi$. A rotation angle of $\pi$ was chosen in each case, but the results are similar for all angles. Each plot corresponds to 500 simulations of the rotation starting from different initial states equally distributed over the Bloch sphere.

\subsection{Noise}
\label{subsec:noise}

Noise is an important source of loss of fidelity in any qubit platform. If unaccounted for, the randomness of noise will lead to gradual loss of quantum information during a computation. While noise mitigation is not the goal of this work, we nonetheless investigate its impact here for completeness. In a quantum-dot-based charge qubit we base our simulations on, charge noise is one of the main noise sources. It arises from fluctuations of charge states that lead to fluctuations of electric field a qubit experiences \citep{Kuhlmann2013}. \\
 Here, we use a simple model where the charge noise is low-frequency and can be assumed to be constant during a single quantum operation \citep{PhysRevB.100.165305}. In practice, this could result from some charge trapped temporarily on one side of the DQD, imparting an electric field gradient, effectively adding an unwanted random bias voltage. Therefore, to calculate the resulting fidelity loss, we average the resulting fidelity from many simulations, each with a random amplitude. The effective Hamiltonian has an additional noise term:

\begin{equation}\label{eq:Hnoise}
\hat{H}_{\mathrm{noise}}(t) = -\frac{1}{2} \,[\epsilon(t)+\delta_{\mathrm{noise}}] \,\sigma_{x} + \frac{1}{2} \,\Delta \,\sigma_{z}  ,
\end{equation}

where $\delta_{\mathrm{noise}}$ is the noise amplitude randomly drawn from a normal distribution with mean $\mu=0$ and standard deviation $\sigma_{\mathrm{noise}}$, which quantifies noise strength.\\

A large number (order of 100) of simulations are run with this randomised noise for some example operations, and the effects of this noise are compared between a square wave, and adjusted pulses accounting for rise time that are the result of this work. The random number generator seed is the same for both cases, so that they experience exactly the same noise and thus can be compared fairly. We find that the effects of charge noise on $R_{\vec{y}}$ and $R_{\vec{z}}$ rotations are not affected by our pulsing method. This is to be expected, as the pulse was not designed with noise in mind. At the very least, we confirm that our proposed pulse is not any worse than an idealised square wave, and further error mitigation techniques can be applied to it, as they would be to a square pulse, without it causing any loss of fidelity, while the problems associated with rise time are solved.\\

However, we find that there is a subset of cases where our pulse sequence does produce a reduction in noise-related errors. When performing an $R_{\vec{x}}$ rotation, it is possible to sub-divide the pulse into further smaller sub-pulses that add up to the total angle of rotation $\theta$. This is only possible when the rise time constraint allows for such a division, as there will exist a minimum angle that you cannot subdivide further.  For the $R_{\vec{z}}$ rotation however, this method doesn't work well, as the angle $\Theta_2$ is always relatively large, even for small total rotation angle $\alpha$. Therefore, attempting to subdivide a larger rotation would result in a very long total operation time, as the total angle that needs to be rotated is no longer (approximately) proportional to $\alpha$. The case for $R_{\vec{y}}$ suffers from similar issues as $R_{\vec{z}}$, therefore one cannot use this optimisation by subdivision to improve resilience against noise. The dependence of total rotated angle (which approximately corresponds to total operation time) on the required rotation angle $\alpha$ is different for $R_{\vec{y}}$ and $R_{\vec{z}}$ rotations, compared to $R_{\vec{x}}$, therefore noise reduction occurs only in the latter. 

\begin{figure}
\centering
\includegraphics[width=0.5\textwidth]{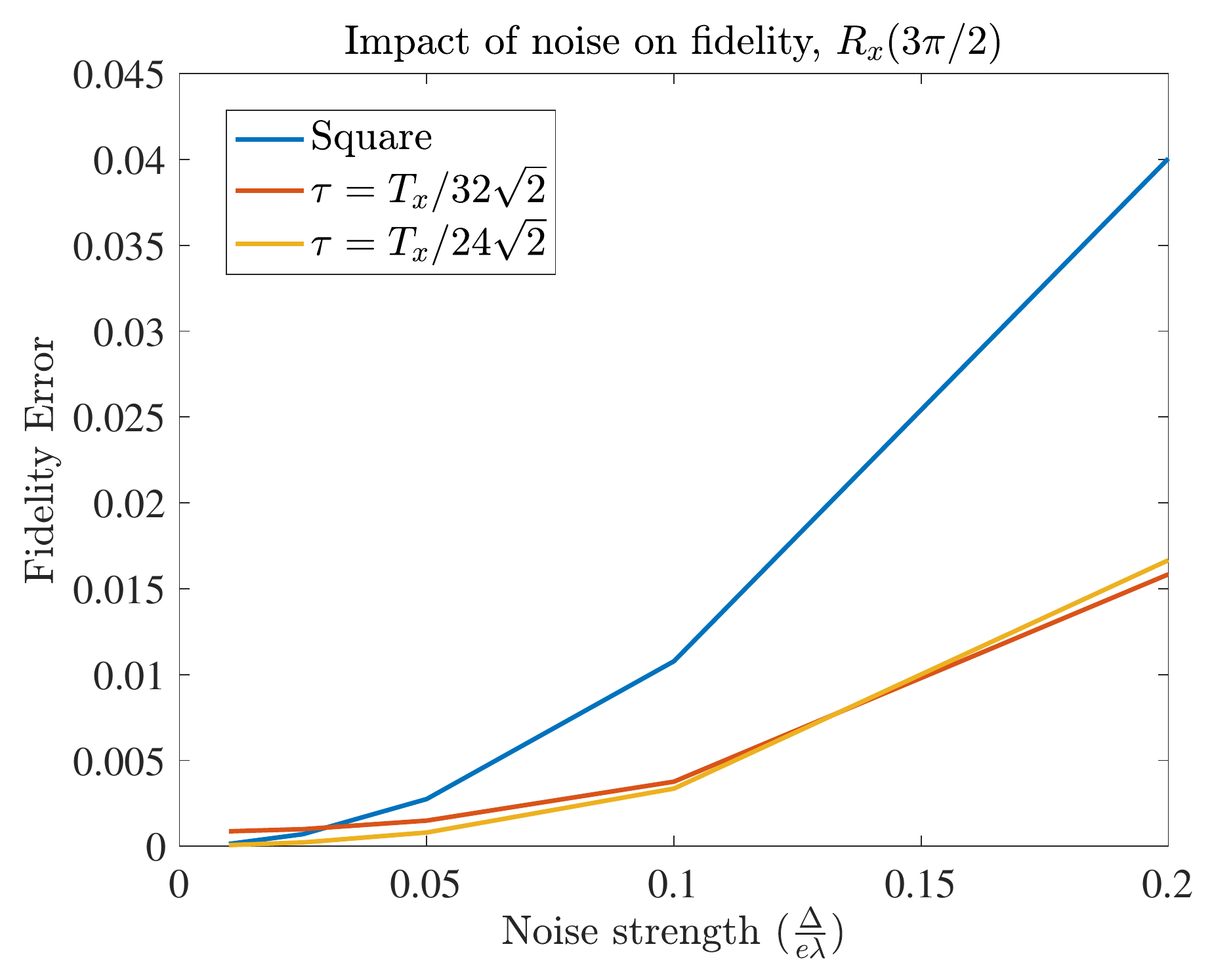}
\caption[Gain in fidelity]{Up to 250 \% gain in fidelity is observed when performing an $R_{x}$ rotation by $\theta=\frac{3}{2}\pi$. Noise strength is expressed in units of reference pulse strength $\sigma_{\mathrm{noise}} = \frac{\Delta}{e \lambda}$, and varied from $0.01 \frac{\Delta}{e \lambda}$ to $0.2 \frac{\Delta}{e \lambda}$.}
\label{fig:noise2}
\end{figure}

An example of noise reduction owing to subdivision into smaller pulses for an  $R_{\vec{x}}$  rotation  is presented in Fig. \ref{fig:noise2}.\\

\begin{figure}
\centering
\includegraphics[width=0.5\textwidth]{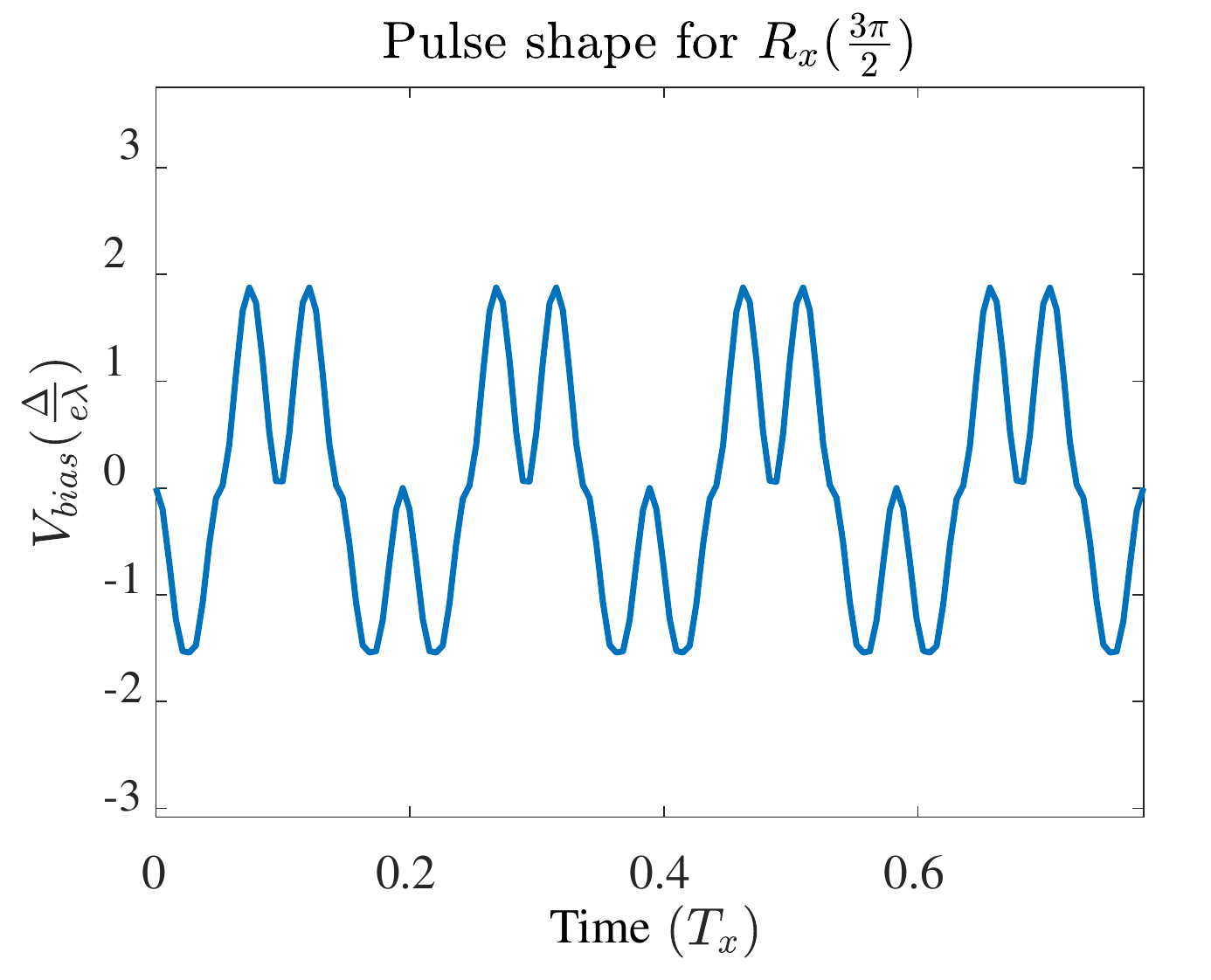}
\caption[Pulse increasing fidelity]{A gain in fidelity is observed when performing an $R_{x}$ rotation by $\theta=\frac{3}{2}\pi$. The resulting pulse shape oscillates from negative to positive multiple times, effectively canceling out some of the noise. Noise strength is expressed in units of reference pulse strength $\sigma_{\mathrm{noise}} = \frac{\Delta}{e \lambda}$, and varied from $0.01 \frac{\Delta}{e \lambda}$ to $0.2 \frac{\Delta}{e \lambda}$.}
\label{fig:noisePulse}
\end{figure}

This beneficial effect of subdividing the pulse can be understood by investigating the pulse sequence that  achieves the rotation. As seen in Fig. \ref{fig:noisePulse}, which shows a pulse shape of a noise-reducing sequence, the oscillating nature of the pulse takes it from being negative to positive frequently. This will average out the influence of noise to a significant degree, while keeping the total operation time close to the one for an ideal square wave. \\

Overall, we conclude that the control techniques presented here are at least as good in resisting noise as using a square wave, and can improve upon it under certain conditions. Therefore, they are suitable to replace the square wave, and to have further noise-reducing methods applied upon them, while they offset any errors due to rise time. The optimised $R_{\vec{x}}$ rotation is able to mitigate charge noise up to almost threefold in the fidelity error (this gain increases with noise strength), given that rise time $\tau$ enables one to perform multiple smaller rotations that add up to a required total angle.

\section{Discussion}\label{sec:discussion}

When experimentally optimising qubit rotations, voltage pulses are usually considered as square while rise and fall times from instrument limitations and other filtering effects due to the finite bandwidth of coaxial cables are neglected. While the voltage is gradually rising to some intended amplitude, the qubit will undergo transient rotations, and will not reach the expected position on the Bloch sphere. These errors accumulate over long operations, leading to poor fidelity. Moreover, applying very sharp pulses of high amplitude, with the intent of performing an $R_z$ rotation, can lead to unwanted energy excitations due to non-adiabacity, causing further fidelity loss ~\citep{landau32, zener32, stueckelberg32}. The control scheme presented here overcomes both problems by explicitly adjusting the pulses for rise time, and by using relatively low pulse amplitudes, making the operations adiabatic. By using a specific amplitude giving us two perpendicular rotation axes, we achieve single-qubit control without the need for strong non-adiabatic pulses, or the requirement for perfectly square ones. The disadvantage of this scheme is operation time. As the pulse amplitude is tied to the energy gap between the first two eigenstates of the DQD, there is little control of the rotation speed, at least in the case of a semiconductor DQD system. However, careful engineering of the DQD allows for the operation time to be tailored or optimized ~\citep{lepage2020}. As long as the system energies can be tuned so that the operation time is much less than qubit coherence time, the benefits of increased operation fidelity will outweigh the cost of increased duration.

In this work, we simulate a semiconductor GaAs-based DQD using finite difference methods (App. ~\ref{appendix:ischeme}). The parameters for our simulations were chosen to be experimentally realistic in terms of energy, time scales and pulse generation. We kept these values general since specific rise/fall times and inter-dot energies will depend on each experimental implementation. Current systems are capable of generating pulses with $\tau=$ 40ps-500ps. Experimental work by Fujisawa \emph{et al.}~\cite{hayashi03, fujisawa04} contain gate pulses with  $\tau \sim 100$ps, with total pulse time of $600$ps and $V_{\mathrm{bias}} = 40 \mu$eV. More recent work achieves at least 40ps pulse resolution with advanced techniques ~\citep{eriksson2020}.

The groups cited above as well as other semiconductor-based quantum dot research ~\citep{taylor2007, petta05} could practically eliminate errors due to rise time and pulse-induced excitations outside of the computational space by using our proposed pulse sequences.

While the semiconductor charge qubit system was used in simulations in this work, our results are easily generalisable to other types of qubits, as long as the Hamiltonian is of a similar form to Eq. \ref{eq:H1}. For example, the same scheme can be used to control a spin qubit by varying the magnetic field $B$ instead of a voltage bias. In this particular case, it is easier to adjust the energy splitting $\Delta = \frac{1}{2} \gamma B$ by applying a strong reference magnetic field. Increasing $\Delta$ will result in faster operation. However, in the charge qubit case, it is achieved by lowering the DQD barrier. This will increase the overlap between eigenstates, decreasing localization and thus readout fidelity. No such issue arises for the spin qubit, overcoming the slower operation time of our framework. Our results can then be directly translated to the spin qubit case, by applying a magnetic field $B^{\prime}$ in some perpendicular direction to $B$.

\section{Conclusions}\label{sec:conclusion}

We have described quantum control of the optimal charge qubits for a double-quantum dot system. We presented pulse sequences for state preparation and arbitrary qubit rotations, and show how to account for the experimental control suffering from finite rise/fall times. Owing to hybridization of the eigenstates in a double-dot system, the spatial wave function of the two lowest energy eigenstates cannot be confined exclusively to the left and the right dot. The optimal qubit was found to be defined in terms of the two lowest energy eigenstates of a zero-bias system. This allowed us to reduce our model to a two-state system.

We show that it is possible to prepare the qubit in such a state when it is initially in the ground state of a DQD. Combining theory and numerical techniques yields an optimal pulse sequence that accomplishes arbitrary single-qubit rotation even with non-zero rise time $\tau$. We demonstrate how our framework results in high fidelity despite $\tau > 0$, while avoiding unwanted excitation to higher energy states. Indeed, we show that square pulses are not only unnecessary, but also undesirable, as the sharp rise can induce unwanted oscillations, while being simple to account for. Since our proposed pulse sequence reduces to sine waves to minimize total pulse duration, it is straightforward to implement experimentally. As our numerical fitting parameters depend only on the energy splitting $\Delta$, the results are easily scalable to any particular system. Our scheme is easily generalizable to other qubit systems with similar Hamiltonians, such as spin qubits.

Additionally, we study a model of charge noise, and find that our pulse scheme is at least as good as using square waves, and it some cases it even significantly reduces errors due to noise. This further justifies using our method as a direct replacement for square waves, as other noise mitigation and error correction techniques can be used on top of it.

Overall, applying our results will lead to increased operation fidelity in many systems, making them viable for practical quantum computing applications.

Our method of accounting for rise/fall times bears resemblance to the GRAPE ( Gradient Ascent Pulse Engineering)  algorithm \cite{rowland2011}, however there are important differences. Our method specifically works to cancel the rise/fall times of assumed profile (sinusoidal in this work, but the method can be used for any shape), resulting in a simple lookup of two parameters $\xi$ and $\Delta T$ depending on required angle of rotation and $\tau$ itself. GRAPE instead is a more general \enquote{black box} technique that tries to optimise a pulse sequence by constructing it from slices of piecewise constant amplitudes, by tuning these amplitudes via gradient ascent methods. This research can also be used to optimize current geometric approaches to pulse shaping ~\citep{edbarnes} by taking rise times into account explicitly.
We find that the method used here is simpler to implement for experimentalists, outputs a waveform composed of sinusoids, which can be described analytically, and is, by design, not limited by the device rise/fall time.

\section{Acknowledgements}
This project has received funding from the European Union’s Horizon 2020 Research and Innovation Programme under the Marie Skłodowska-Curie grant agreement SAWTRAIN No. 642688. This work was supported by the Project for Developing Innovation Systems of the Ministry of Education,Culture, Sports, Science and Technology (MEXT), Japan. A.A.L. acknowledges support from Hitachi via Grant No. RG94632, and from EPSRC (Engineering and Physical Sciences Research Council) via Award No. 1948709.

\section{Data availability}
The data that support the findings of this study are available from the corresponding authors on reasonable request.

\section{Competing interests}
The authors declare that there are no competing interests.

\section{Author contributions}
A.L. and H.V.L. designed the code to simulate the wave function evolution, developed the theory in this work and ran all simulations. A.L. and H.V.L. wrote the manuscript with the help of all authors. K.Z. helped design the optimized pulse sequence. T.F. provided experimental parameters for realistic simulations. C.H.W.B. supervised the project. All authors discussed the simulation results.

\appendix

\section{Readout}
\label{sec:readout}

For completeness, we discuss a potential procedure for the readout process. In experimental setups, it is the probability of finding the electron in one of the dots which is measured rather than the qubit superposition weighting coefficients.  We can express both qubits defined in Sec.~\ref{sec:methods} in terms of their right and left dots parts:

\begin{eqnarray}
\label{eq:LRzero}\psi_0(x)=\braket{x|0} =  f_{0L}(x) + f_{0R}(x) \\
\label{eq:LRone}\psi_1(x)=\braket{x|1} = f_{1L}(x) + f_{1R}(x)
\end{eqnarray}

Because the qubits $\ket{0}$ and $\ket{1}$ are orthogonal, we have:

\begin{align}
\begin{split}
\label{eq:readoutA}
0 = \int \psi^*_0(x) \psi_1(x) dx = \int f^*_{0L}(x) f_{1L}(x) dx + \\
\int f^*_{0L}(x) f_{1R}(x) dx + \int f^*_{0R}(x) f_{1L}(x) dx +\\
 \int f^*_{0R}(x) f_{1R}(x) dx = \int f^*_{0L}(x) f_{1L}(x) dx +\\
  \int f^*_{0R}(x) f_{1R}(x) dx.
\end{split}
\end{align}

The qubits are mirror images of each other, such that $\braket{x|0}$ has the same spatial distribution in the left (right) dot as $\braket{x|1}$ has in the right (left) one. We also know that there is some non-zero overlap, unless the DQD barrier is completely separating the dots. Therefore Eq.~\ref{eq:readoutA} implies that :

\begin{equation}\label{eq:readoutB}
\int f^*_{0R}(x) f_{1R}(x) dx = -\int f^*_{0L}(x) f_{1L}(x) dx  = \eta.
\end{equation}

Any arbitrary state can be written as a linear combination of the two qubits right and left dot components

\begin{align}
\begin{split}
\label{eq:readoutC}
\psi(x) = \alpha \psi_0(x) + \beta \psi_1(x) = \\
\alpha \Big(f_{0L}(x) + f_{0R}(x)\Big) + \beta \Big(f_{1L}(x) + f_{1R}(x)\Big),
\end{split}
\end{align}

The probability $P_{R}$ of finding the particle in the right dot is then:

\begin{align}
\begin{split}
\label{eq:readoutD}
P_{R} = \int_0^\infty \psi^*(x) \psi(x) dx = 
\int_0^\infty \Big( \alpha^* f^*_{0R}(x) + \\
\beta^* f^*_{1R}(x) \Big)\,\Big( \alpha f_{0R}(x) +
\beta f_{1R}(x) \Big) dx.
\end{split}
\end{align}

Using Eq.~\ref{eq:readoutB}, this reduces to:

\begin{align}
\begin{split}
\label{eq:readoutE}
P_{R} = |\alpha|^{2}  \int_0^\infty f^*_{0R}(x) f_{0R}(x) dx  + \\
|\beta|^{2} \int_0^\infty f^*_{1R}(x) f_{1R}(x) dx + \eta (\alpha^* \beta + \alpha \beta^* )   =\\ |\alpha|^{2}  P_{0R} + |\beta|^{2}  P_{1R} + 2 \eta \mathcal{Re}(\alpha^* \beta ) ,
\end{split}
\end{align}

where the integrals $P_{0R}$ and $P_{1R}$ can be obtained initialising the qubit in the $\psi_0(x)$ or $\psi_1(x)$ state, respectively, and measuring the probability of finding it in the right dot.  Combining Eq.~\ref{eq:readoutE} with the normalisation condition for $\psi(x)$, we obtain an equation relating $|\beta|$ to the probability $P_{R}$ of finding the particle in the right dot, up to an error term proportional to $\eta$, which quantifies the uncertainty of determining whether the qubit is in the left or right side of the DQD:

\begin{equation}\label{eq:readoutG}
|\beta|^{2} = \frac{P_{R} - P_{0R}}{P_{1R} - P_{0R}} + \delta.
\end{equation}
A similar expression exists for $|\alpha|^{2}$, with $P_{L}$ being the probability of finding the particle in the left dot :

\begin{equation}\label{eq:readoutGalpha}
|\alpha|^{2} = \frac{P_{L} - P_{0R}}{P_{1R} - P_{0R}} - \delta,
\end{equation}

where $\delta = 2\eta \frac{\mathcal{Re}(\alpha^* \beta )}{P_{0R}-P_{1R}}$ is the effective error. Since $P_{1R} \approx 1$, $P_{0R} \approx 0$, we can estimate the maximum readout error, which would occur for a maximally entangled state:
\begin{equation}
    \mid \delta \mid \lessapprox \eta.
\end{equation}
For the parameters used in this paper, $ \mid \delta \mid \leq 8 \cdot 10^{-4}$.
This magnitude of readout error is not very significant compared to other sources of errors in a quantum computation ~\citep{li2019, huang2019}, such as two-qubit gates, relaxation, or dephasing, especially since it's only applied once as the final step. Additionally, it was shown ~\citep{lepage2020} that in a similar situation, adiabatically increasing the inter-dot barrier of the DQD preserves coherence, while greatly reducing this type of \enquote{overlap} error -this technique should be used when possible if the readout error is noticable.
Alternatively, as this error is a result of lack of knowledge of $\mathcal{Re}(\alpha^* \beta )$, a full state tomography could be performed to eliminate it completely (assuming that errors of operations associated with the tomography do not outweigh the readout error).
Therefore, we conclude that measurement of the charge distribution is a  viable way of reading out the qubit in our scheme.

\section{Two-site localised state model and DQD potential}\label{appendix:2siteLSM}

Within the two-state model, one has to solve the time dependent Schr\"odinger equation with the effective Hamiltonian $\hat{H}_\mathrm{eff}$ defined as
    
\begin{equation}\label{eq:Heff2}
\hat{H}_\mathrm{eff}(t) = -\frac{1}{2} \,\epsilon(t) \,\sigma_{x} + \frac{1}{2} \,\Delta \,\sigma_{z}  + \frac{1}{2}(E_{\textup{B}}+E_{\textup{AB}}).
\end{equation}

Here $E_{\textup{B}}$ and $E_{\textup{AB}}$ are the energies of the bonding and antibonding states of the DQD system, i.e. the two lowest energy states, at $\epsilon$ = 0 whereas $\Delta$ is the 'hybridisation energy' between the two localised states. At zero detuning, the bonding state $\psi^\textup{B}(x)$ is symmetric, while the antibonding state $\psi^\textup{AB}(x)$ is antisymmetric. Therefore, their equal superpositions produce maximally localised left/right states:
\begin{equation}
    \psi^L(x)=\frac{1}{\sqrt{2}}(\psi^\textup{B}(x) +\psi^\textup{AB}(x)),
\end{equation}
\begin{equation}
    \psi^R(x)=\frac{1}{\sqrt{2}}(\psi^\textup{B}(x) -\psi^\textup{AB}(x)).
\end{equation}

The linear detuning breaks the left/right symmetry, however as it is expressed by the Pauli $\sigma_x$ matrix in the Hamiltonian, it doesn't make the system leave the $\psi^\textup{B}(x) / \psi^\textup{AB}(x)$ two-state basis (if done adiabatically), resulting simply in a coordinate rotation of the Bloch sphere. Therefore, we can still think in terms of the left/right localised wave functions even at non-zero detuning, and varying $\epsilon$ is a viable way of performing single qubit rotations.
The two-site localised state model describes a DQD well. The effective potential in an experimental DQD system can be found using density functional theory~\cite{owen15,stopa96} and will be a complex function of all three spatial coordinates $x,y,z$.  By careful design, the dynamics in two of the directions $y$ and $z$ can be confined to the lowest energy subbands so that only the potential in the $x$ direction, $V_{\textup{DQD}}(x,t)$ needs be considered.  For example in a GaAs/AlGaAs heterostructure, the $z$ direction is the growth direction and modulation doping can be used to create a triangular quantum well in that direction with subband energies two orders of magnitude larger than either $\epsilon$ or $\Delta$.  In the $y$ direction, parabolic confinement with energies an order of magnitude larger than $\epsilon$ or $\Delta$ can be produced either by etching~\cite{Ferrus11}, fabricating a thin gate wrapping the conducting channel~\cite{hisamoto90,voisin14} or using split-gates~\cite{vanwees88}.  In order to create a DQD potential in the $x$ direction, gates~\cite{fujisawa00,gardelis03,lim09,mason04} or etching~\cite{wei13,Ferrus11} can also be used.

The aim is to create a potential $V_{\textup{DQD}}(x,t)$ that has two minima separated by a tunnel barrier.  A convenient potential that has this property and is defined by three parameters $A$, $B$ and $\sigma$ is given by

\begin{equation}\label{eq:vdqd}
V_{\textup{DQD}}(x) = A x^2 + B \exp\left(\frac{-x^2}{2\sigma}\right)
\end{equation}

This form for $V_{\textup{DQD}}$ allows us to control both the depth of the dots and the barrier between them directly, by varying the harmonic confinement $A$, barrier height $B$, and barrier width $\sigma$.
This potential will obey the two-site localised state model. For a specific set of parameters, this static potential will define a value for $\Delta$ which is the energy difference between the bonding ground state $E_{\textup{B}}$ and the antibonding first excited state $E_{\textup{AB}}$.  Detuning is introduced by adding a linear Stark shift of the form

\begin{equation}\label{eq:vbias}
V_\mathrm{linear} (x) = V_\mathrm{bias} \frac{x}{2 w}.
\end{equation}

\begin{figure}
\centering
\includegraphics[width=0.7\columnwidth]{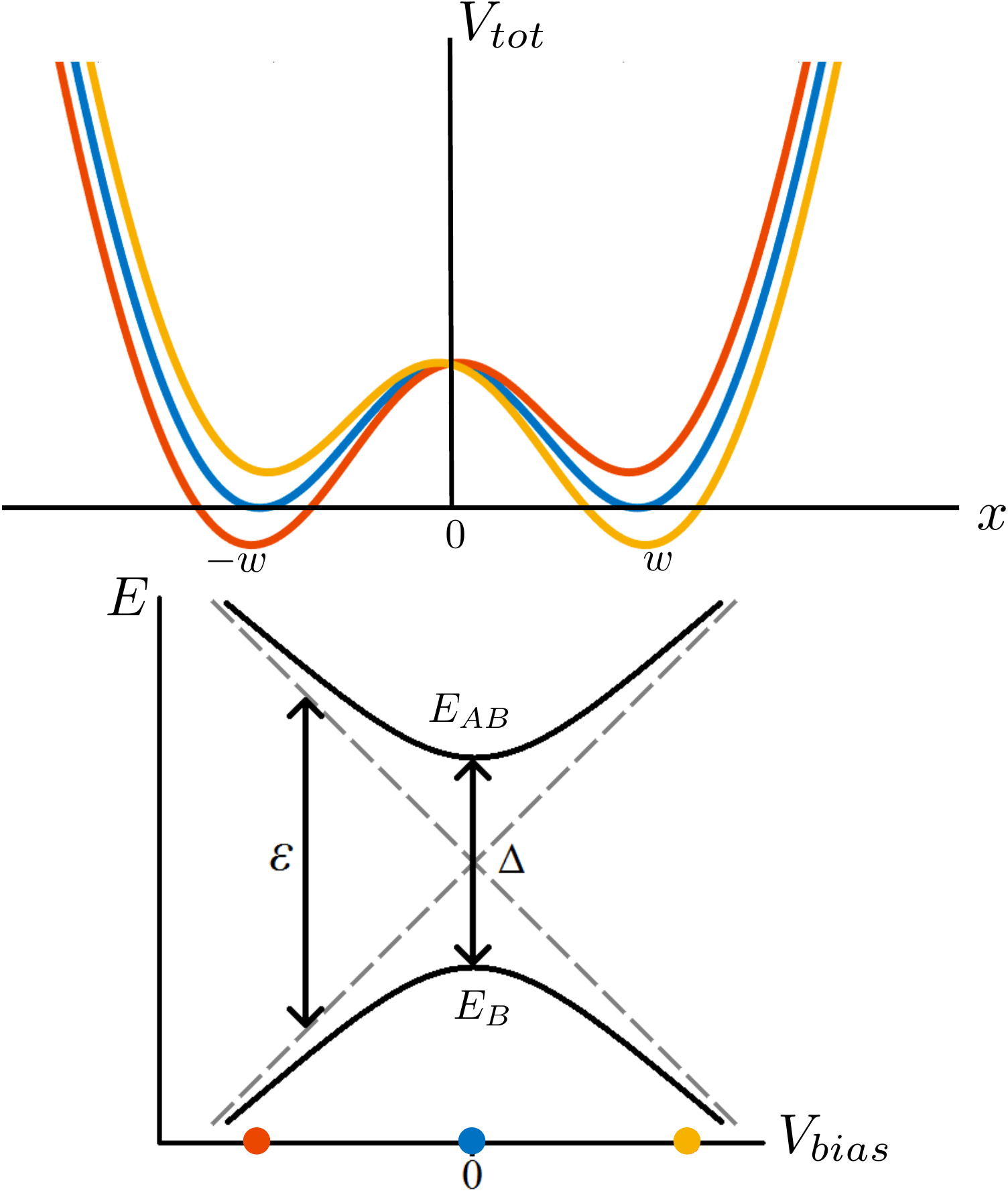}
\caption{(a) The DQD potential $V_\mathrm{tot}$ at zero (blue), lowest (red) and highest (orange) detuning values. (b) Energies $E$ of the bonding ($E_{\textup{B}}$) and anti-bonding ($E_{AB}$) eigenstates.  The coloured dots mark potential shapes from part (a).}
\label{fig:potentiel} 
\end{figure}

Here, $w$ is half the width of the DQD. By comparing the dependences of $E_{\textup{B}}$ and $E_{\textup{AB}}$ on $V_\mathrm{bias}$  with the expected dependences from two-site Hamiltonian we can define the detuning parameter for $V_{\textup{DQD}}$ through a linear relation $\epsilon = e \lambda  V_\mathrm{bias}$ with $\lambda$ being constant.  We find this linear relationship holds with an accuracy of one part in $10^6$ across the range of required values of $\epsilon$ for single-qubit operations. The total potential is $V_\mathrm{tot}(x) = V_\mathrm{\textup{DQD}}(x) + V_\mathrm{linear}(x)$ and Fig.~\ref{fig:potentiel}a shows this potential at three different detunings.

The DQD dynamics under time-dependent detuning will be given by the TDSE

\begin{equation}\label{eq:TDSEreal}
\hat{H}(x,t) \psi(x,t) = i \hbar \frac{\partial}{\partial t}\psi(x,t)
\end{equation}

with

\begin{equation}\label{eq:realHamiltonian}
\hat{H}(x,t) = - \frac{\hbar^2}{2 m^*} \frac{\partial^2}{\partial x^2} + V_{\textup{DQD}} (x) + V_{\mathrm{bias}} (x,t).
\end{equation}

Time dependence is included in Eq.~\ref{eq:TDSEreal} by varying the potential slope with time: $V_\mathrm{bias}(t)$.  An example plot of the energies of the two lowest instantaneous solutions (the bonding and antibonding states) as function of $V_\mathrm{bias}$ is shown in Fig.~\ref{fig:potentiel}b.

Analytic solutions to the TDSE in Eq.~\ref{eq:realHamiltonian} can only be found in special cases.  In this paper we solve Eq.~\ref{eq:TDSEreal} numerically using a GPU-accelerated version of the staggered-leapfrog method~\cite{askar78, owen12} (see App.~\ref{appendix:ischeme}).

Throughout the paper we avoid using specific numerical values to keep our results general. However, here we give the actual values used for reproducability. We've use a total DQD length of $460$  $\textnormal{nm}$, with parameter values:  $w = 230 $ nm, $A = 1.276$ meV $\textnormal{nm}^{-1}$, $B = 4.08$ meV so that $\mathit{\Delta} = 11.7 \mu\textnormal{eV}$, and the linear coefficient $\lambda = 0.421$.  We have also tested various non-symmetric potentials with the two dots having different sizes, but in all the cases the general conclusions were the same as for the symmetric potential of Eq.~\ref{eq:vdqd}.

\section{Rotation scheme derivation}\label{appendix:AnalyticalRotations}
We find a fast and simple general rotation scheme based on creating two perpendicular axes $\vec{x}^{\prime}$ and $\vec{z}^{\prime}$, by setting the detuning $\epsilon=\pm \Delta$. Then we observe that one should be able to perform a rotation around an axis at $\frac{\pi}{4}$ w.r.t the two axes above, which would be $\vec{x}$ and $\vec{z}$. This is achieved by rotating by some angle $\Theta_1$ around the first axis, then by $\Theta_2$ around the second one, and finally by $\Theta_1$ around the first one again.

We will find the relationship between $\Theta_1$, $\Theta_2$, and the net angle rotated around $\vec{x}$ or $\vec{z}$ named $\alpha$, by analytically comparing the rotation matrix elements with the straightforward 
$R_{\vec{x}}$ and $R_{\vec{z}}$ rotations. 

Looking at $R_{\vec{x}}$ first:

\begin{equation}
R_{\vec{x}}(\alpha) =
    \begin{pmatrix}
    \cos{\frac{\alpha}{2}} & -i \sin{\frac{\alpha}{2}}   \\
    -i \sin{\frac{\alpha}{2}} & \cos{\frac{\alpha}{2}} \\
    \end{pmatrix}.
    \label{eq:RotationXStandard}
\end{equation}
In our scheme,

\begin{equation}
R_{\vec{x'}}(\alpha) = R_{\vec{y}}\left(\frac{-\pi}{4}\right) R_{\vec{x}}(\alpha) R_{\vec{y}}\left(\frac{\pi}{4}\right),
\end{equation}

\begin{equation}
R_{\vec{z'}}(\alpha) = R_{\vec{y}}\left(\frac{-\pi}{4}\right) R_{\vec{z}}(\alpha) R_{\vec{y}}\left(\frac{\pi}{4}\right),
\end{equation}
and we need the following to always hold:

\begin{equation}
\label{eq:App_Rx}
R_{\vec{x}}(\alpha) = R_{\vec{x'}}(\Theta_1) R_{\vec{z'}}(\Theta_2) R_{\vec{x'}}(\Theta_1).
\end{equation}
Comparing the (1,1) matrix elements:

\begin{align}
\begin{split}
\label{eq:App_RxMatrix11}
\cos{\alpha} =\cos{\frac{\Theta_2}{2}} (2\cos^2{\frac{\Theta_1}{2}} -1)  \\
-j\sqrt{2}( \frac{1}{2} \sin{\frac{\Theta_2}{2}} + \cos{\frac{\Theta_1}{2}} \cos{\frac{\Theta_2}{2}} \sin{\frac{\Theta_1}{2}}).
\end{split}
\end{align}

Since the imaginary part on the LHS is zero, we have:
\begin{equation}
 ( \frac{1}{2} \sin{\frac{\Theta_2}{2}} + \cos{\frac{\Theta_1}{2}} \cos{\frac{\Theta_2}{2}} \sin{\frac{\Theta_1}{2}}) = 0.
\end{equation}
Solving the above allows us to find $\Theta_2$ in terms of $\Theta_1$:
\begin{equation}
\Theta_2 = 2\arctan\left({\sin{\Theta_1}}\right).
\end{equation} 
Now coming back to the real part of Eq. \ref{eq:App_RxMatrix11} and substituting for $\Theta_2$, we have:
\begin{equation}
\cos{\left( \arctan{\left(\sin{\Theta_1}\right)} \right)}\cos{\Theta_1} = \cos{\frac{\alpha}{2}},
\end{equation} 
which gives
\begin{equation}
\Theta_1 = \arccos\left( \frac{\sqrt{2}\cos{\frac{\alpha}{2}}}{\sqrt{\cos{\frac{\alpha}{2}}^2 +1}}\right).
\end{equation}

The above satisfies Eq. \ref{eq:App_Rx} for all matrix elements, and is therefore equivalent. It allows us to find a three pulse train that performs the $R_{\vec{x}}(\alpha)$ rotation by an arbitrary angle $\alpha$.
We repeat the above procedure for $R_{\vec{z}}$
to find the following:
\begin{equation}
\Theta_2 = -2\arctan\left({\sin{\Theta_1}}\right) +2\pi,
\end{equation} 

\begin{equation}
\Theta_1 = \arccos\left( \frac{\sqrt{2}\cos{\frac{\alpha}{2}}}{\sqrt{\cos{\frac{\alpha}{2}}^2 +1}}\right).
\end{equation}

Therefore, we can perform arbitrary rotations around
$\vec{z}$ and $\vec{x}$ this way. However, this scheme is unable to perform the $R_{\vec{y}}$ rotation, which is achieved differently, as described in \ref{sec:general rotation scheme}.

\section{Iteration method}\label{appendix:ischeme}

The system is modelled using an explicit iterative scheme for the one-dimensional time-dependent Schr\"odinger equation (TDSE) with an arbitrary potential \emph{V(x,t)}:

\begin{equation}\label{eq:TDSE}
i \hbar \frac{\partial \psi(x, t)}{\partial t} = H\psi = \left[\frac{-\hbar^{2}}{2 m} \frac{\partial^{2}}{\partial x^{2}} + V(x, t)\right] \psi(x, t)
\end{equation}

where \emph{m} is the effective mass. The scheme, which is based on the finite difference method, was described in details by Maestri \emph{et al.} for two particles in one dimension~\cite{maestri00} and we adapt it to a single particle. The wave function is evaluated on a spatially discretized grid and at successive, equally separated intervals of time $\Delta t$:

\begin{equation}\label{eq:psi}
\psi(x, t) = \psi(m \Delta x, k \Delta t) \equiv \psi^{k}_{m},
\end{equation}

with $m, k$ integer.  The spatial part of the method is derived using Taylor expansion of the wave function:

\begin{equation}\label{eq:xderiv}
\frac{\partial^{2} \psi}{\partial x^{2}} \simeq \frac{\psi(x+\Delta x) - 2 \psi(x) + \psi(x-\Delta x)}{\Delta x^{2}}.
\end{equation}

Therefore, using Eqs.~(\ref{eq:psi}) and~(\ref{eq:xderiv}), the right hand side of Eq.~(\ref{eq:TDSE}) transforms into

\begin{equation}\label{eq:TDSEdiscrete}
H \psi = \Bigg[\frac{-\hbar^{2}}{2 m}\bigg( \frac{\psi_{m+1} - 2 \psi_{m} + \psi_{m-1} }{\Delta x^{2}} \bigg)\, +  \,V_{m}\Bigg] \psi_{m}.
\end{equation}

The derivative on the left hand side of Eq.~(\ref{eq:TDSE}) is calculated by writing the exact solution of TDSE and then taking the difference between the $(k\!+\!1)^{th}$ and $(k\!-\!1)^{th}$ time steps, as suggested by Askar and Cakmak~\cite{askar78}:

\begin{equation}\label{eq:psinext}
\psi^{k+1}_{m} = e^{-i \Delta t H / \hbar} \psi^{k}_{m} \simeq \left(1 - \frac{i \Delta t H}{\hbar}\right) \psi^{k}_{m},
\end{equation}

\begin{equation}\label{eq:psidiff}
\psi^{k+1}_{m}\, - \,\psi^{k-1}_{m} = (e^{-i \Delta t H / \hbar} - e^{i \Delta t H / \hbar}) \psi^{k}_{m} \simeq - \frac{2 i \Delta t H}{\hbar} \psi^{k}_{m}.
\end{equation}

To improve the accuracy, we follow Visscher's staggered-time method~\cite{visscher91} and write the wave vector in terms of its real and imaginary parts: $\psi^{k}_{m} = u^{k}_{m} + i v^{k}_{m}$.  After inserting the Hamiltonian from Eq.~(\ref{eq:TDSEdiscrete}) into Eq.~(\ref{eq:psidiff}) and rearranging the terms, we obtain a pair of simultaneous equations, which are iterated over time:

\begin{eqnarray}\label{eq:uv}
u^{k+1}_{m} = u^{k-1}_{m} + 
 \Big( 2a_{x}+b\,V^{k}_{m} \Big) v^{k}_{m} - \\ 
 a_{x} (v^{k}_{m+1}\!+\!v^{k}_{m-1}) ,\\
v^{k+1}_{m} = v^{k-1}_{m} - 
 \Big( 2a_{x}+b\,V^{k}_{m} \Big) u^{k}_{m} - \\
 a_{x} (u^{k}_{m+1}\!+\!u^{k}_{m-1}) ,
\end{eqnarray}

where $a_{x}=\frac{\hbar \Delta t}{m \Delta x^{2}}$ and $b=\frac{2 \Delta t}{\hbar}$.  Also, the real and imaginary parts are calculated at slightly shifted times: $u^{k} \equiv u(t),\; v^{k} \equiv v(t+\Delta t/2).$

The method above is stable as long as the following criterion is satisfied:

\begin{equation}
\Delta t \leq \frac{\hbar}{E_{\mathrm{max}}},
\end{equation}

with $E_{\mathrm{max}}$ being the largest eigenvalue of the discretised Hamiltonian~\cite{leforestier91}.  Furthermore, small errors due to finite computational accuracy do not accumulate with iterations and the total electron probability $\sum_{\mathrm{all\,} m} |\psi^{k}_{m}|^2$ is preserved over time, showing no significant deviations from unity.

\section{Finding optimal adjustment parameters accounting for rise time $\tau$ by gradient ascent}\label{appendix:pseudocode}

The MATLAB code for finding the optimal adjustment parameters accounting for rise time $\tau$ for single qubit control is available on request from the corresponding author.    The time-dependent evolution is relegated to the GPU-accelerated staggered-leapfrog code described and referenced in the main work.

\bibliography{main.bib}

\end{document}